# THE EFFECT OF AIR EXPOSURE ON SEY AND SURFACE COMPOSITION OF LASER TREATED COPPER APPLIED IN ACCELERATORS

Jie Wang (王洁)[1, a], Taaj Sian[2,3,4], Reza Valizadeh[2,3], Yong Wang（王勇）[5], Sheng Wang（王盛）[1, a]

[1]School of Energy and Power Engineering, Xi'an Jiaotong University, Xi'an, Shaanxi, China
[2]ASTeC, STFC Daresbury Laboratory, Warrington, UK
[3]Cockcroft Institute, Warrington, UK
[4]University of Manchester, Manchester, UK
[5]National Synchrotron Radiation Laboratory, University of Science and Technology of China, Hefei, AnHui, China

*[a] Corresponding author. Jie Wang, E-mail address: wangjie1@xjtu.edu.cn    Sheng Wang, E-mail address: shengwang@xjtu.edu.cn*

## *Abstract*

In this paper, the effect of air exposure and acetone cleaning on the secondary electron yield (SEY) of laser treated copper were investigated and discussed. The SEY of laser treated copper samples after acetone cleaning and air exposure were measured. After 10 to 21 months air exposure, the maximum SEY increased by about 11%~20%. Furthermore, changes of carbon, copper, and oxygen chemistry states from sample surface after acetone and air exposure were studied by means of XPS. The XPS results show that the concentration of Cu decreased by about 61%~70% after acetone cleaning and decreased by a further 36%~70% after 10 to 21 months air exposure. Moreover, the XPS measurement reflects the significant increase of the total percentages of C and O after acetone cleaning, and further increase after 10 to 21 months air exposure. The increase of SEY is explained in terms of the introduction of impurities, such as oxide, carbide etc.

# 1 Introduction

The electron cloud is a key problem in particle accelerators, which can cause the degradation of beam quality and limit the performance of particle accelerators with high energy, high intensity, high luminosity and long beam lifetime[1-8]. The methods of mitigating electron cloud effect have been studied by many researchers from several organizations, such as CERN[9, 10], STFC[11, 12], NSRL[13], KEK[7, 14], INFN[15], etc. C. Yin Vallgren et al. found that a complete suppression of e-cloud can be achieved by amorphous carbon coating of liners[9]. Yusuke Suetsugu et al. tested the effectiveness of antechambers and TiN coating for the inhibition of electron cloud[14]. Valizadeh et al. proposed laser engineered surface technology to obtain low SEY surface[11, 12]. After laser treatment, the SEY of stainless steel, copper, and aluminum can be reduced to less than 1. Comparing to other methods for obtaining low SEY surface or materials, laser engineered surface technology has the advantages of no need for vacuum or clean room environment, highly reproducible, inexpensive etc. Taking these advantages into account, laser engineered surface technology is currently an ideal method to obtain low SEY surface.

However, laser engineered samples will be exposed to air inevitably. Therefore, the influence of the air exposure on the SEY and surface composition of such samples is worth studying. Copper is used as the inner surface of beam screen [16, 17] due to its qualities for maintaining good vacuum environment, and handling heat load induced by synchrotron radiation etc. Therefore, copper plates and copper foils were used as the samples.

In the following, the SEY measurement parameters and XPS test results of laser treated copper samples are presented. The effects of the air exposure on the SEY and surface composition of laser-engineered copper samples are also discussed.

# 2 Experimental

## 2.1 SEY measurement method

The SEY values obtained based on the principle shown in Eq. (1). Here $I_p$, $I_s$ and If are the current of incident electrons, sample-to-ground current and Faraday cup-to-ground current, respectively. The schematic of SEY measurement equipment is shown in Figure 1. The biased voltage on the sample holder is set to -18 V, with the distance between the faraday cup and the sample of 1 mm.

$$\delta_{SEY} = \frac{I_f}{I_p} = \frac{I_f}{(I_f + I_s)} \tag{1}$$

The SEY test equipment consists of an EGPS-2B electron gun, Faraday cup, sample holder, two Keithley 6485 picoammeter, a source of X-ray, electron energy analyzer, electron detector, sample transfer frame, and power supply.

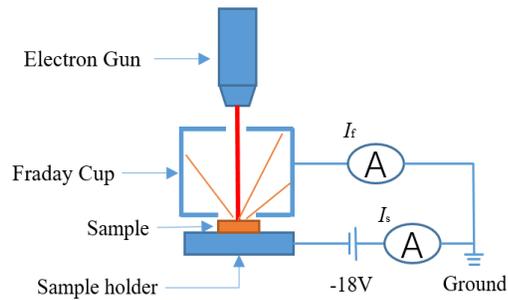

FIG. 1. (Color online) The schematic of the SEY measurement equipment.

## 2.2 SEY test parameters

The electron dose per unit was $7.6\times10^{-8}$ C mm$^{-2}$. The electron gun （Kimball, ELG-2/EGPS-2） scans over an energy range of 50 eV to 1000 eV. In addition, the background pressure of the chamber before the SEY test was $(4-8)\times10^{-10}$ Torr and $(2-8)\times10^{-9}$ Torr during the SEY test. The test temperature was 300 K.

## 2.3  Laser parameters for treating copper samples

The rolled and electro-polished oxygen-free copper (Cu) samples were commercially available with high purity of 99.999%. The samples with the size of 20 mm ×20 mm were degreased prior to laser engineering. The samples were cleaned with acetone and then were treated in air and at room temperature by pulses of a Nd:YAG laser. This laser was operated at various repetition rates (see Table 1). The diameter of the laser spot with a Gaussian intensity profile on each samples was set at 25 μm for λ=355 nm and 15 μm for λ=1064 nm. The beam diameter, the scanning speed and the pulse repetition determine the pulse number reaching the samples. Different laser parameters used for this study are shown in Table 1.

TABLE I. Laser parameters for laser treated copper samples.

| Sample | Pulse duration /ps | Spot /μm | Scan speed /mms$^{-1}$ | Wave length /nm | Pitch spacing /μm | Repetition /kHz | Power /W |
|---|---|---|---|---|---|---|---|
| #1 | 5 | 15 | 30 | 1064 | 5 | 100 | 5 |
| #2 | 2000 | 25 | 120 | 355 | 5 | 40 | 5 |
| #3 | 2000 | 25 | 90 | 355 | 5 | 40 | 5 |

## 3  Results and discussion

## 3.1 SEY results

The SEY of Samples is shown in Figure 2 (a) - (c) as a function of primary electron energy, $\delta(E_p)$ for the three different scan speeds. It shows that for the as-received sample the SEY remains below 1. Initially, the SEY rises when the primary electron energies is below 250 eV. The rate of increase reduces towards the higher primary electron energy.

The maximum SEY of as-received, after acetone and 10 months air exposure laser treated oxygen-free copper sample #1 were 0.92, 1.03 and 1.15, respectively, as shown in Figure 2(a). And the corresponding incident electron energies were 682 eV, 682 eV, and 632 eV, as shown in Table 2. After acetone cleaning, the maximum SEY of as-received laser treated copper sample #1 increased by 0.11, about 12%. After 10 months air exposure, the maximum SEY increased by further another 12%, compared to that of after acetone cleaning.

For sample #2, the maximum SEY of as-received and 20 months air exposure were 1.04 and 1.20, respectively, as shown in Figure 2(b). The corresponding incident electron energies were around 982 eV, as shown in Table 2. After 20 months air exposure, the maximum SEY of as-received laser treated copper sample #2 increased by about 16%.

For sample #3, the maximum SEYs of as-received and 21 months air exposure were 0.99 and 1.19, respectively, as shown in Figure 2(c). Furthermore, the corresponding incident electron energies were around 982 eV, as shown in Table 2. After 21 months air exposure, the maximum SEY of as-received laser treated copper sample #3 increased by 0.16, about 15%.

In order to analyze the reason for SEY increase, XPS was used to characterize the surface composition of laser treated copper under different conditions, such as acetone cleaning and air exposure.

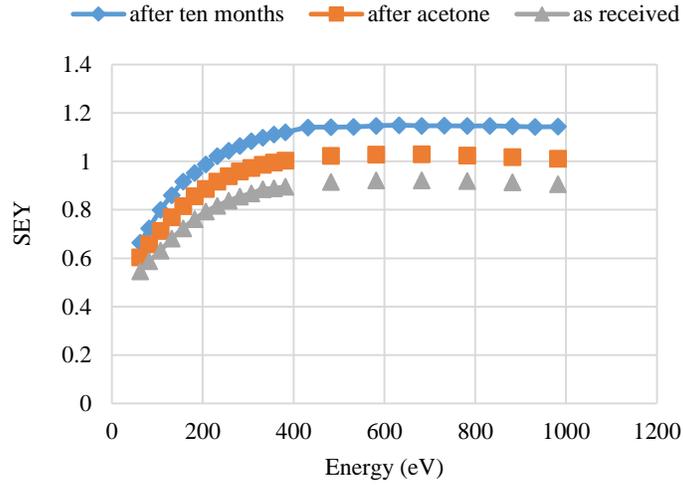

(a) sample #1

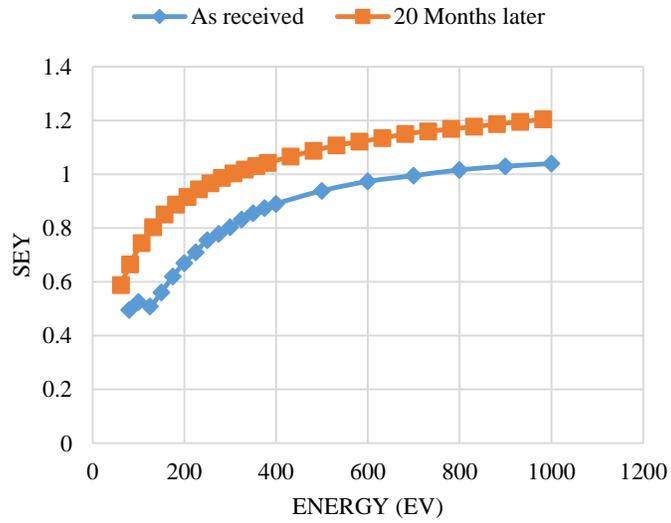

(b) sample #2

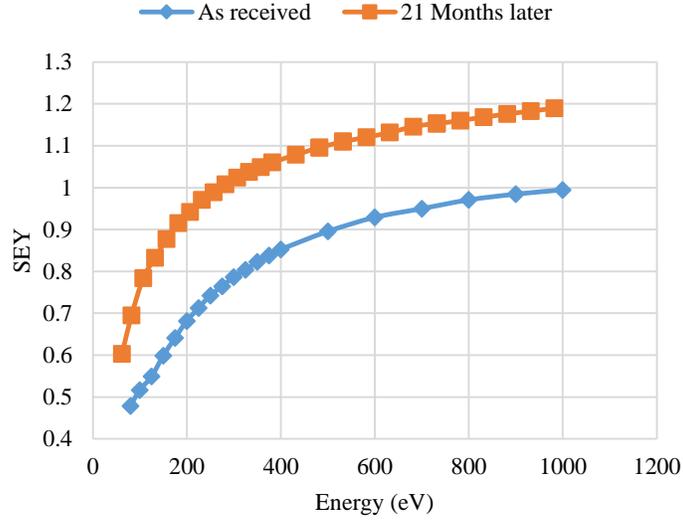

(c) sample #3

FIG. 2. (Color online) SEYs of sample #1, sample #2 and sample #3 under different conditions as a function of primary electron energy with the same electron dose of $7.6\times10^{-8}$ C mm$^{-2}$.

TABLE II. $\delta_{max}$ and $E_{max}$ for as-received, after acetone and after air exposure laser treated Cu samples #1, #2 and #3.

| Sample | SEY parameters | As received | After acetone | After air exposure |
|---|---|---|---|---|
| Cu plate #1 | $\delta_{max}$ | 0.92 | 1.03 | 1.15 (10 months later) |
|  | $E_{max}$ (eV) | 682 | 682 | 632 |
| Cu foil #2 | $\delta_{max}$ | 1.04 | - | 1.20 (20 months later) |
|  | $E_{max}$ (eV) | 982 | - | 982 |
| Cu foil #3 | $\delta_{max}$ | 0.99 | - | 1.19 (21 months later) |
|  | $E_{max}$ (eV) | 982 | - | 982 |

## 3.2 XPS analysis of laser treated copper samples

The surface composition and chemical state on as-received, acetone cleaning and air exposure laser treated Cu samples was studied using XPS.

Figure 3 represents the XPS survey spectra of copper samples #1 for the three different conditions. The copper signal at the surface is represented with two groups of peaks Cu 2p3/2 at 933 eV and Cu 2p1/2 at 953 eV due to spin–orbit coupling. The spectra exhibit creation of stoichiometric CuO at the surface. The shake-up satellite structure at a binding energy of 943 eV indicated that it consists mainly of cupric oxides. The presence of this satellite structure has been attributed to charge transfer transitions from the ligands ($O^{2-}$ ions) into the unfilled valence level of the $Cu^{2+}$ ion[18]. Compared to the Cu/Cu2O peak, the broader peak shape of the cupric species is due to coupling between unpaired electrons[18, 19]. Oxygen O1s and carbon C1s are represented by peaks at 532 eV and 284 eV respectively and their atomic percentages for each sample at different condition are presented in Table 3. The concentration of Cu under different conditions, as-received, after acetone cleaning, and 10 months air exposure were 78.7%, 29.1%, and 14.6%, respectively. The results show that the concentration of Cu decreased about 63% after acetone cleaning and 81% after 10 months air exposure. While, the concentration of Cu from as received laser treated copper foil samples were 23.2% and 18.8%, respectively. It indicates that the percentage of carbon-containing compounds and oxygenated chemicals from the surface of as received laser treated copper foils are much higher than that of as received laser treated copper plate sample, as shown in Table 3.

TABLE III. The concentration comparison of laser treated copper under different conditions, as-received, after acetone cleaning and ten months air exposure laser.

| Conditioning | Elements | Cu 2p At% | C 1s At% | O 1s At% | N 1s At% |
|---|---|---|---|---|---|
| Sample #1 | as-received | 78.7 | 2.8 | 18.4 | 0 |
| | after acetone cleaning | 29.1 | 17.3 | 53.6 | 0 |
| | 10 months air exposure | 14.6 | 38.7 | 46.8 | 0 |
| Sample #2 | as-received | 23.3 | 28.5 | 48.2 | 0 |
| | after acetone cleaning | 7.4 | 46.5 | 46.1 | 0 |
| | 20 months air exposure | 4.7 | 60.9 | 26.6 | 7.8 |
| Sample #3 | after acetone cleaning | 18.8 | 38.7 | 42.5 | 0 |
| | 21 months air exposure | 5.7 | 56.6 | 31.4 | 6.3 |

The XPS spectra of the laser treated copper sample #1, under the conditions of acetone cleaning and air exposure are shown in Figure 3. The ratio of C, O and Cu elements for the surface of the as-received laser treated Cu were 2.8%, 18.4%, and 78.7%, respectively. After cleaning in acetone, the ratio of C, O and Cu elements on the sample #1 surface were 17.7%, 53.6%, and 29.1%, respectively. After air exposure, these ratios were 38.7 %, 46.8%, and 14.6% respectively, as shown in Table 3. In contrast with the surface composition of laser treated copper, the proportion of C and O increased significantly by about 6 times and 3 times respectively after acetone cleaning, probably due to the residual of acetone. After 10 months air exposure, the proportion of C and O increased even further because of the further oxidation of Cu on the surface of the sample and the introduction of adventitious impurities.

The increase of impurities containing C and O on sample surface after acetone cleaning may cause the increase of SEY of laser treated copper sample #1. Moreover, the further increase of SEY after air exposure may be caused by the introduction of impurities such as carbon, hydroxides, and oxy-organics on laser treated copper surface.

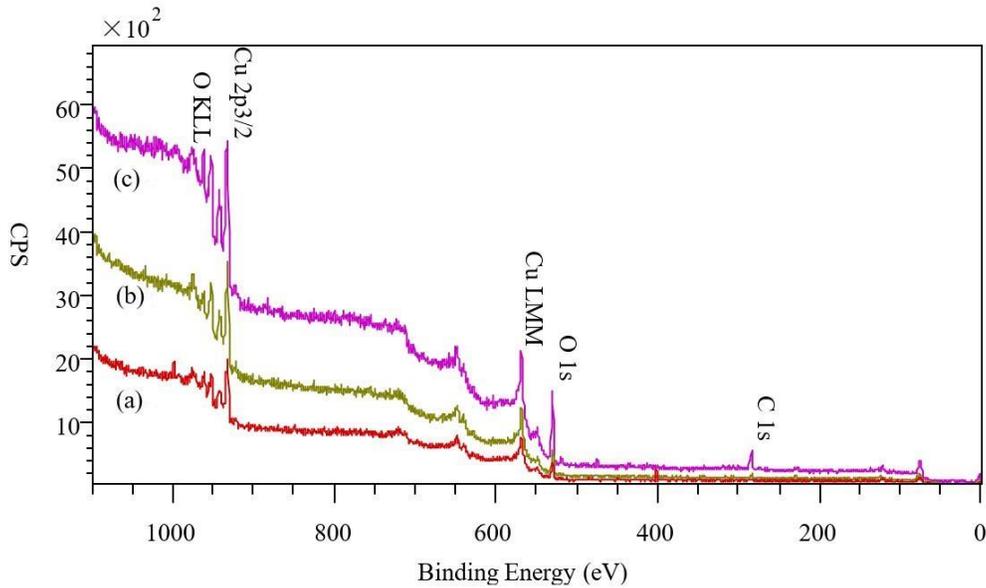

FIG. 3. (Color online) The XPS spectra of (a) as-received (Red), (b) after acetone cleaning (Green), and (c) ten months air exposure (purple) from laser treated copper sample #1.

3.2.1 Curve fitting of Cu 2p spectrum

According to the XPS spectrums of as-received, cleaned in acetone and 10 months air exposure sample #1, it shows that the concentration of Cu under three different conditions are 78.7%, 29.1%, and 14.6%, respectively. This means that the content of Cu from laser treated copper sample #1 decreased by 63% after acetone cleaning. Then it reduced further to 14.6% after 10 months air exposure.

As shown in Figure 4, it is quite clear that there is a sharp peak here and the lost structure and the broader peak that can be associated with a two-plus state. It suggests that the percentage of cuprous (Cu I) and cupric (Cu II) of the as-received laser treated copper samples #1 were 18.7% and 81.3%, respectively. After acetone cleaning, they were 17.8% and 82.2%. In addition, they were 17.9% and 82.1% after 10 months air exposure. Consequently, the percentage of Cu (II) after 10 months air exposure was very close to the one of as received laser treated copper sample #1, which indicate that there is no obvious further oxidation of Cu mental.

According to the XPS curve fitting of Cu 2p spectrum from sample #2, it indicates that the percentage of Cu (II) after 20 months air exposure decreased slightly by 4% comparing with that from as received laser treated copper sample #2, shown in Figure 5. Therefore, it indicates that there is slightly further oxidation of Cu mental after 20 months air exposure.

Figure 6 shows that there is no obvious peak shift for Cu 2p3/2 under three different conditions i.e. as received, after acetone, 21 months air exposure wrapped in aluminium. The percentage of Cu (II) from sample #3 is nearly 100%, which is the reason for no obvious peak shift. This means the chemical state of Cu from as received copper sample #3 is cupric.

Finally, it can be concluded that the chemical state of Cu is mainly cupric after laser treatment. The percentage of Cu (II) varied between 82%~99%. These results show that the percentage of Cu (I) and Cu (II) are similar, possibly slightly further oxidation under different conditions, which means that the oxidation states of the copper element after laser treatment are essentially stable. Therefore, the increase of SEY is mainly caused by the introduction of impurities, such as oxide, carbide etc. Next, the XPS spectra of C and O from laser treated Cu sample will be analyzed.

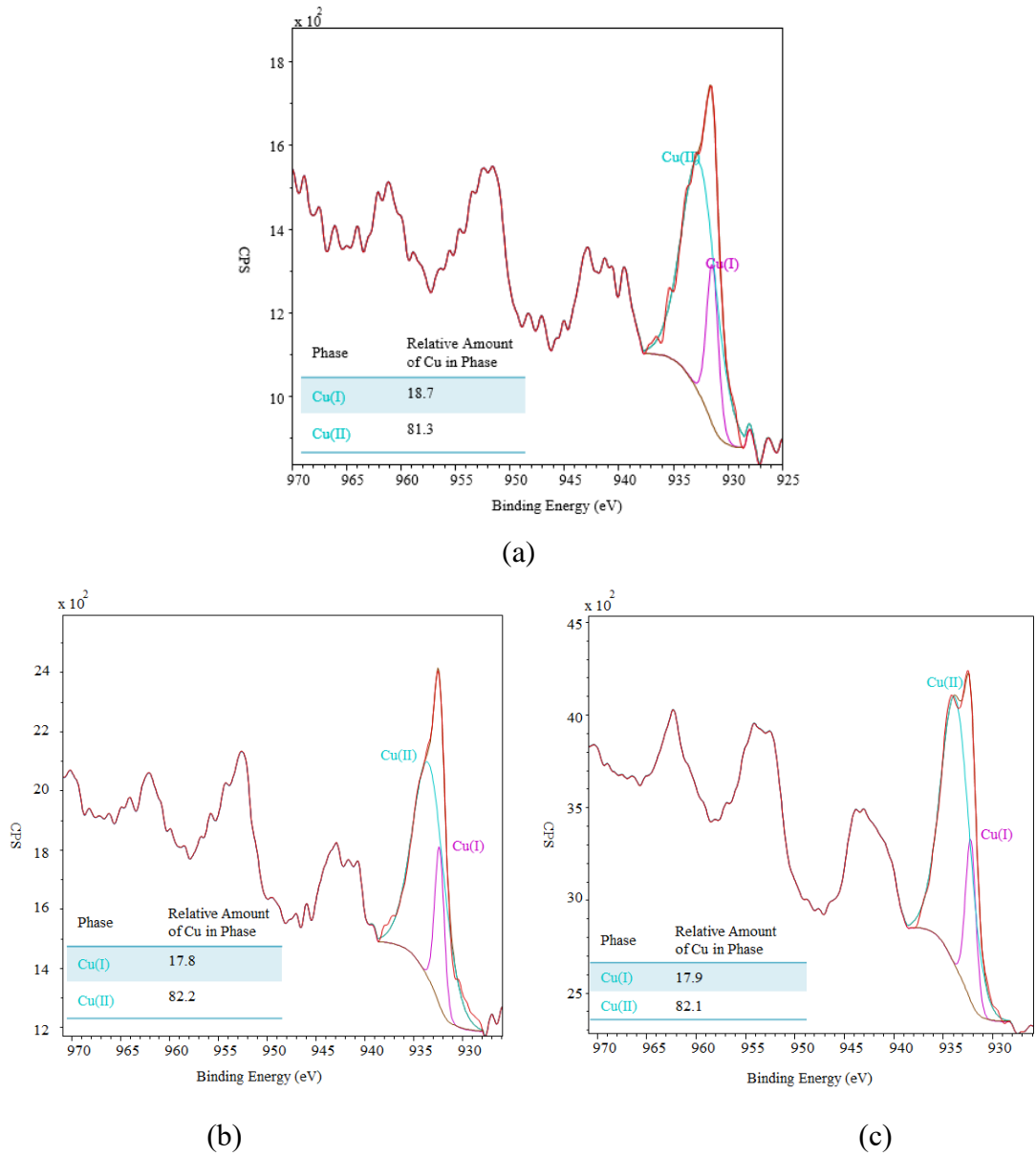

FIG. 4. The XPS spectra of Cu 2p from laser treated Cu sample #1, (a) as received, (b) after acetone and (c) 10 months later.

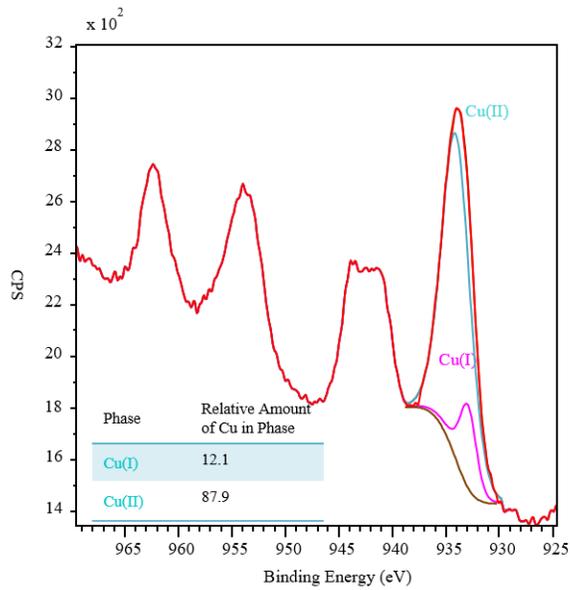

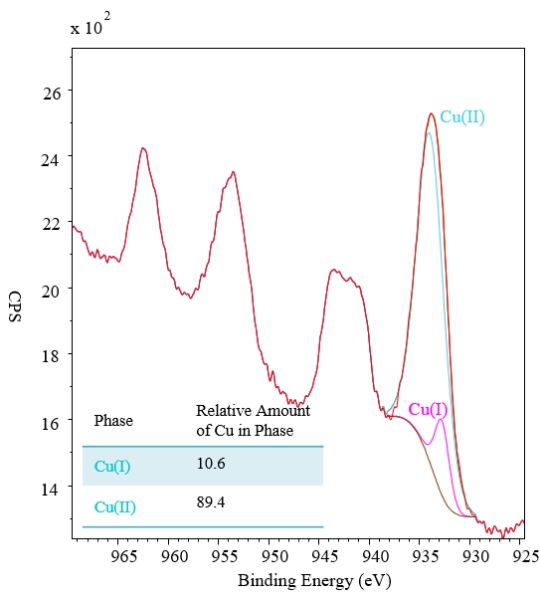

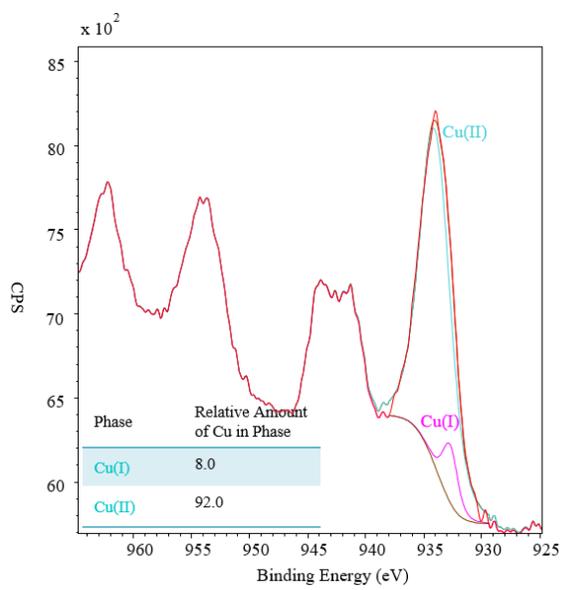

FIG. 5. The XPS spectra of Cu from laser treated Cu sample #2, (a) as received, (b) after acetone and (c) 20 months later.

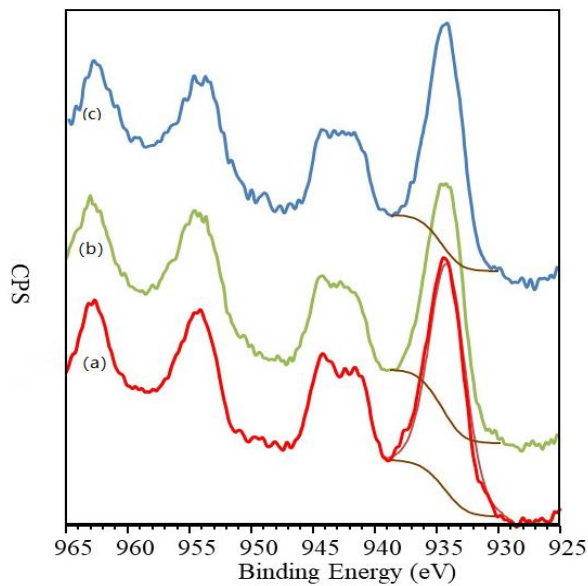

FIG. 6. The XPS spectra of Cu 2p from laser treated Cu sample #3, (a) red-as received, (b) green-after acetone and (c) blue-21 months later.

3.2.2 Curve fitting of C1s spectrum

The binding energy of C–C and C–H were close to 284.8 eV. As shown in Figure 7, curve fitting of C1s spectra of laser treated copper samples under different conditions was performed by using the Gaussian (70%)–Lorentzian (30%) peak-shapes which was defined in CasaXPS as GL(30) and Shirley background subtraction. In addition, C-O-H, O-C=O, and C=O refer to alcohol, ketones, esters or acids, respectively. The XPS spectrum of C from laser treated Cu sample #1 shows that the total percentage of C-C and C-H from as-received, after acetone and 10 months air exposure Cu sample, are 63.98%, 69.9%, and 72.1%, respectively. The total percentage of C-O-H, O-C=O, and C=O are 36.02%, 30.1%, and 27.9% under different conditions. The comparison shows that the total percentage of C-C and C-H from as-received laser treated Cu sample #1 increases slightly after acetone cleaning and 10 months air exposure. Carbon is the adventitious contaminants for laser treated copper sample and the percentage of C–C and C–H increased obviously after 10 to 20 months air exposure.

As shown in Figure 8, C–C and C–H is the main component of adventitious carbon for laser treated Cu sample #2, with a percentage of 64%~91%. After 20 months air exposure, the total percentage of alcohol, ketones, esters or acids was 8%. In short, it manifests that carbon-free environment is important for reducing the percentage of adventitious carbon.

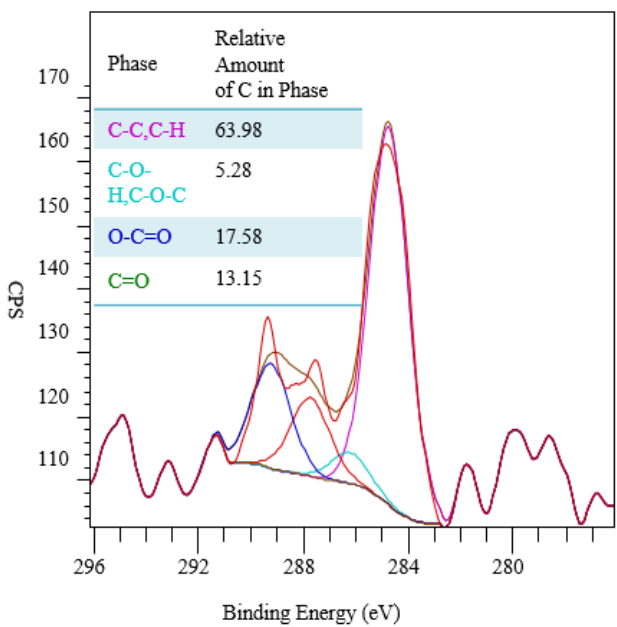

(a)

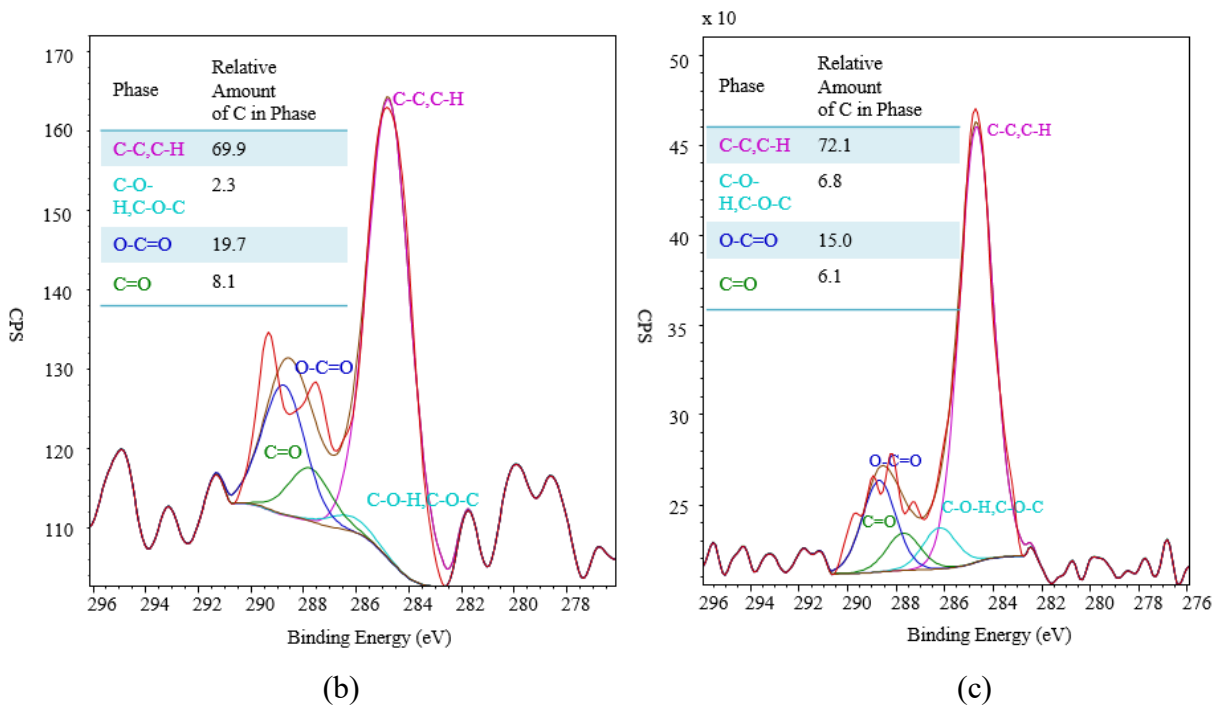

FIG. 7. The XPS spectra of C 1s from laser treated Cu sample #1: (a) as received, (b) after acetone and (c) 10 months later.

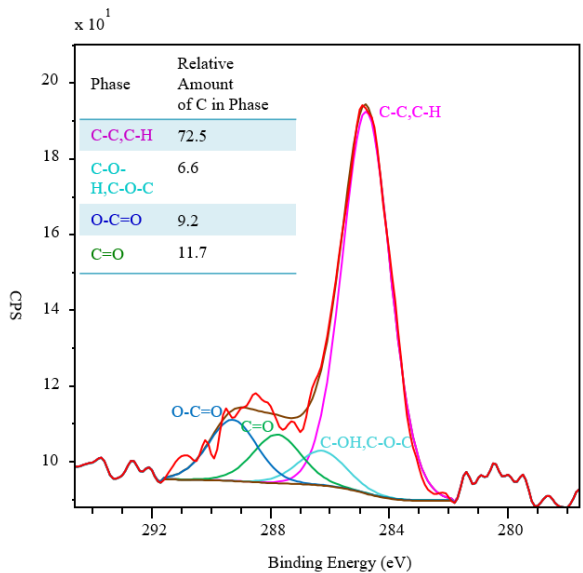

(a)

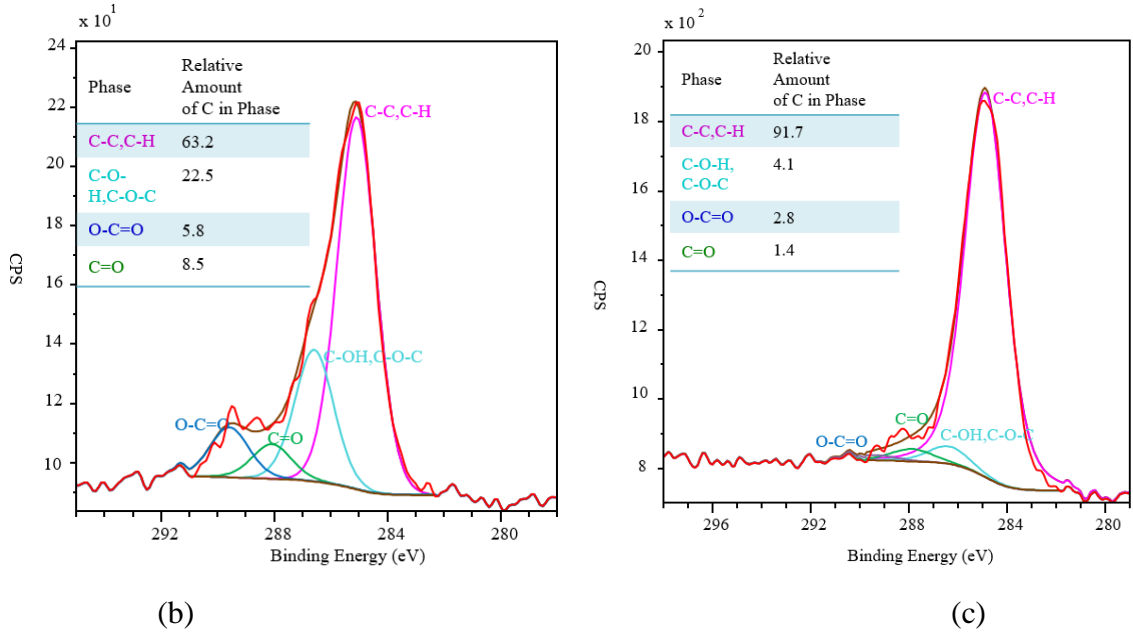

(b)                                (c)

FIG. 8. The XPS spectra of C from laser treated Cu sample #2, (a) as received, (b) after acetone and (c) 20 months later.

3.2.3 Curve fitting of O1s spectrum

In Figure 9, the XPS spectra of O 1s shows that the percentage of copper lattice oxide from as-received, after acetone and 10 months air exposure Cu sample #1 are 47.44%, 41.20%, and 31.60%, respectively and, the total percentage of hydroxides are 49.69%, 37.6%, and 66% under these three different conditions for sample #1. Moreover, for sample #2, the relative contributions of lattice oxide peak under three different conditions are 46.2%, 44.3%, and 44.0%, respectively as shown in Figure 10. Therefore, lattice oxide and hydroxides are the main components of the total oxygen signal and the percentage of water and organic O from both samples varied between 2%~10%. It suggests that the amount of water increases gradually with the increase of air exposure time. In brief, it manifests that oxygen free environment is important for reducing the percentage of adventitious hydroxides.

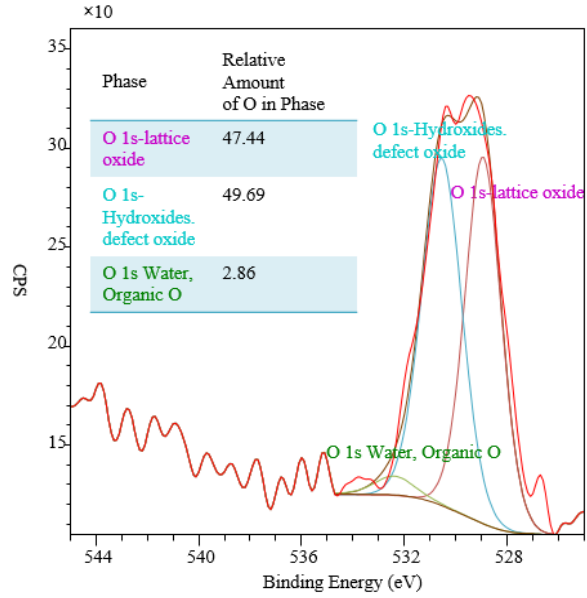

(a)

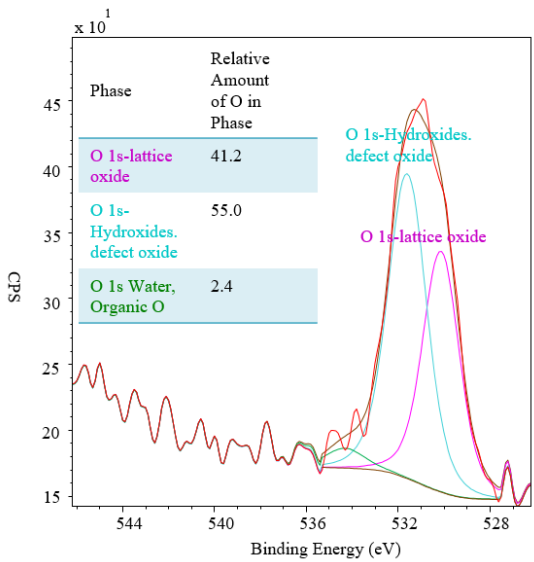

(b)

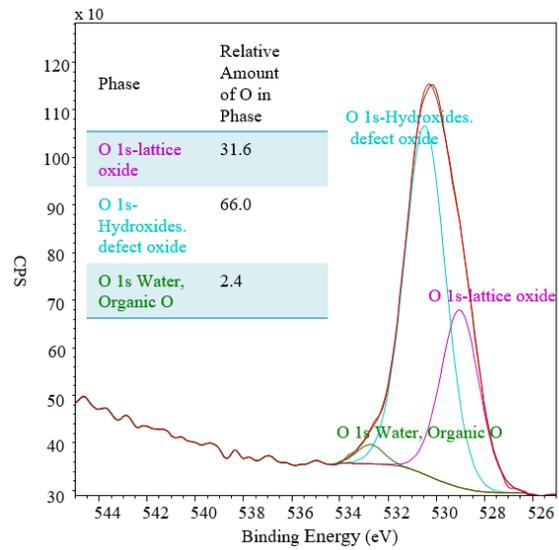

(c)

FIG. 9. The O 1s spectra of the laser treated Cu sample #1, from bottom to top: (a) as received, (b) after acetone and (c) 10 months later.

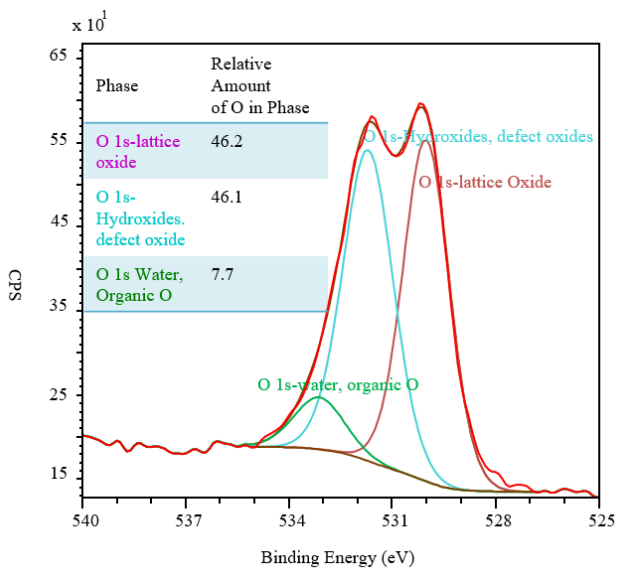

(a)

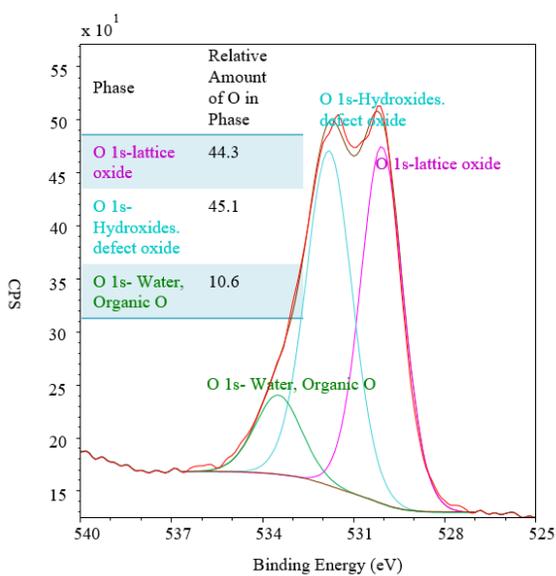

(b)

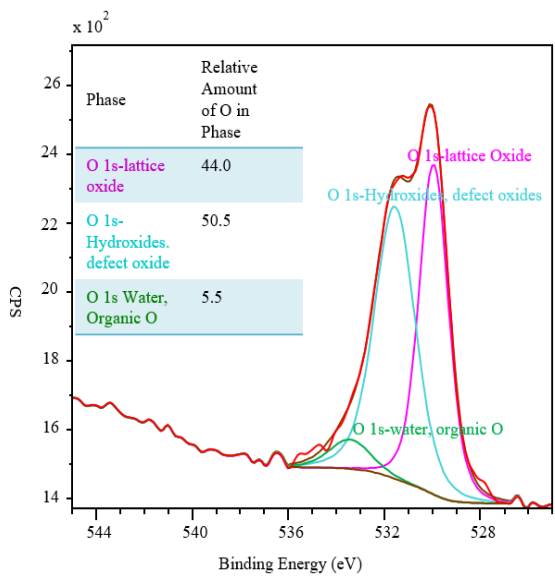

(c)

FIG. 10. The XPS spectra of O from laser treated Cu sample #2, (a) as received, (b) after acetone and (c) 20 months later.

## 4  Summary and conclusions

The effects of acetone cleaning and air exposure on the SEY of laser treated copper samples were investigated and discussed in this paper. The following conclusions were obtained.

(A) After 10 to 21 months air exposure, it indicates that the SEY of laser treated copper samples increases from δmax of 0.92, 1.04, and 0.99 to 1.15, 1.20, and 1.19, respectively. The maximum SEY of laser treated copper increased by 11%~20% after 10 to 21 months.

(B) The XPS results show that the concentration of Cu decreased about 61%~70% after acetone cleaning and further decreased by 36%~70% after 10 to 21 months air exposure due to residual trace of acetone.

(C) The XPS measurements reflect that the percentage of C significantly increased about 15%~19% after acetone cleaning, and further increased about 21%~26% after 10 to 21 months air exposure.

(D) The curve fitting of Cu 2p spectrum showed that chemical state of Cu is mainly cupric with a percentage of 82%~99% after laser treatment. And the curve fitting of C 1s spectrum indicated that C–C and C–H is the main component of adventitious carbon, with a percentage of 64%~91%. In addition, the curve fitting of O 1s spectrum manifested that lattice oxide and hydroxides are the main components of O, with a percentage of 31%~47% and 45%~50%, respectively.

(E) The increase of the SEY is explained in terms of the introduction of impurities, such as oxide, carbide etc. It manifests that carbon and oxygen free environment is important for reducing the percentage of adventitious hydroxides, C-C, and C-H.

# ACKNOWLEDGMENTS

This work was supported by the key project of Intergovernmental International Scientific and Technological Innovation Cooperation in China under Grant No.2016YFE0128900, the National Natural Science Foundation of China under Grant No.11775166, No. 11475166, No.11575214 and No.11205155.